\documentclass[preprint,showpacs,preprintnumbers,amsmath,amssymb]{revtex4}

\usepackage{graphicx}% Include figure files
\usepackage{dcolumn}% Align table columns on decimal point
\usepackage{bm}% bold math

\begin{document}

%\preprint{APS/123-QED}

\title{ Peripheral fragmentation of relativistic $^{11}$B nuclei in photoemulsion}
\author{D.~A.~Artemenkov}
   \affiliation{Joint Insitute for Nuclear Research, Dubna, Russia} 
 \author{V.~Bradnova}
   \affiliation{Joint Insitute for Nuclear Research, Dubna, Russia} 
\author{M.~Karabov\'a}
   \affiliation{University of  P. J.\u Saf\'arik University, Ko\u sice, Slovak Republic} 
\author{A.~D.~Kovalenko}
   \affiliation{Joint Insitute for Nuclear Research, Dubna, Russia}  
 \author{A.~I.~Malakhov}
   \affiliation{Joint Insitute for Nuclear Research, Dubna, Russia} 
\author{G.~I.~Orlova}
   \affiliation{Lebedev Institute of Physics, Russian Academy of Sciences, Moscow, Russia} 
\author{P.~A.~Rukoyatkin}
   \affiliation{Joint Insitute for Nuclear Research, Dubna, Russia}  
\author{J.~Vrl\'akova}
   \affiliation{University of  P. J.\u Saf\'arik University, Ko\u sice, Slovak Republic} 
\author{S.~Vok\'al}
   \affiliation{Joint Insitute for Nuclear Research, Dubna, Russia} 
\author{A.~Vok\'alov\'a}
   \affiliation{Joint Insitute for Nuclear Research, Dubna, Russia}  
 \author{P.~I.~Zarubin}
     \email{zarubin@lhe.jinr.ru}    
     \homepage{http://becquerel.jinr.ru}
   \affiliation{Joint Insitute for Nuclear Research, Dubna, Russia} 

\date{\today}% It is always \today, today,

\begin{abstract}
\indent Experimental results on peripheral fragmentation of relativistic $^{11}$B nuclei are
 presented. In the experiment the emulsions exposured to $^{11}$B beam with momentum 2.75 A~GeV/c
 at the JINR Nuclotron are used. The relative probability of various fragmentation channels  for
 nucleus breakups (class A) and more violent peripheral interactions (class B) have been determined. For
 classes under investigations the sum of the fragment charges in narrow forward cone is equal
 to the projectile charge, but in the events of class A there are no secondary particles and in the
 events of class B there are. In both classes the main channels is $^{11}$B$\rightarrow$2He+X: 62\%
 and 50\%, corresponding.\par 
\indent  The main channel $^{11}$B$\rightarrow$2$\cdot$(Z$_{fr}$=2)+(Z$_{fr}$=1) was investigated
in details. Momentum measurements of single-charged fragments have been done to determine
 number of p, d and t in the channel. This way it was found that the ratio N$_p$ : N$_d$ : N$_t$ is about
 1:1:1 for $^{11}$B nuclei dissociation and about 15:5:1 for peripheral interactions of  $^{11}$B
 nuclei.
\par
\end{abstract}
  %    {PACS-key}{21.45.+v} \and
   %   {PACS-key}{23.60+e} \and
    %  {PACS-key}{25.10.+s}  
 \pacs{21.45.+v,~23.60+e,~25.10.+s}

\maketitle
\section{\label{sec:level1}Introduction}
\indent 
The study of peripheral interactions  of light odd-even nuclei $^{7}$Li and $^{11}$B in nuclear track emulsion can provide 
a ground for including tritons as clusters into the general pattern of multiple fragmentation of heavier nuclei \cite{web}. 
It is established that in the \lq\lq white\rq\rq ~stars produced by relativistic 
$^{7}$Li nuclei in most peripheral collisions, the $^{7}$Li$^*\rightarrow\alpha+t$ channel constitutes as high as 50\% 
\cite{AdamovichPh04,Adamovich04}. 
In this way a dominant role  a triton as a cluster with lowest separation energy (2.47 MeV) has been revealed
 for the case of relativistic
$^{7}$Li nuclei.  The present $^{11}$B experiment is a logical continuation of the $^{7}$Li study aiming 
to establish the probabilities of 
the low threshold channels, namely, $^7Li+\alpha$ (8.67 MeV), $t+2\alpha$ (11.22 MeV),
and $^{10}Be+p$ (11.23 MeV). In particular, it will allow one to verify whether exists a  correlation
between a channel threshold value, fragment number and composition, and the corresponding propability.

\par
\indent 
The $^{11}$B nucleus is a daughter one in 
the $\beta$ decay of its mirror  nucleus $^{11}$C having a very similar level structure. As application, the  present  study will provide
a comparison ground to explore in future the $^3$He role as a cluster in low threshold breakups  of a $^{11}$C nucleus: $^7Be+\alpha$ (7.54 MeV),
$^3He+2\alpha$ (9.22 MeV), and $^{10}Be+p$ (8.69 MeV) and to evalute a Coulomb effects in a few-body fragmentation.
\par

\section{\label{sec:level2}Experiment}

\begin{figure*}
    \includegraphics[width=15cm]{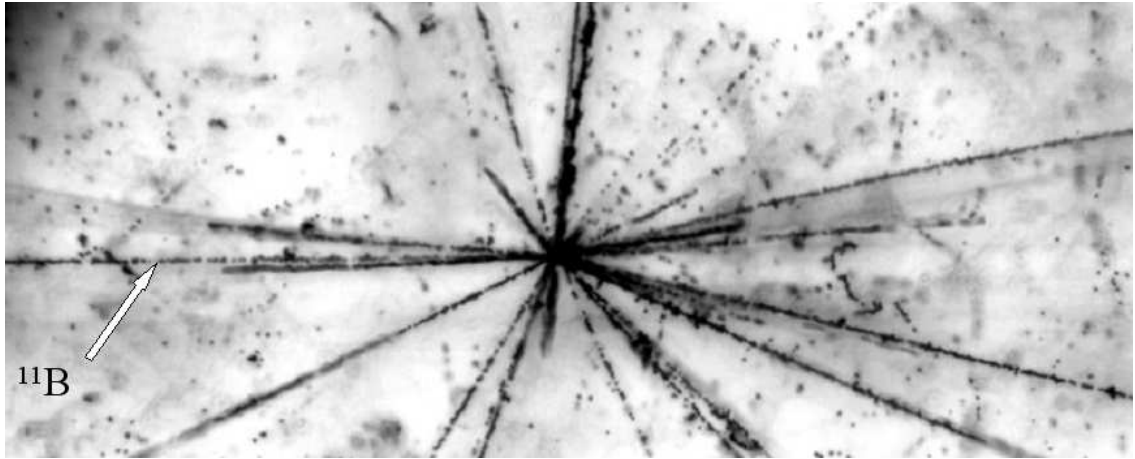}
    \caption{\label{fig:111} An event of central interaction of relativistic $^{11}$B nucleus with
 emulsion nucleus.}
    \end{figure*}

\indent  A stack of BR-2 photoemulsion layers, the dimensions and the thickness of which being 10$\times$20 cm$^2$
and 600 $\mu$m, repectively, was exposured to a beam of  $^{11}$B nuclei accelerated to a momentum 2.75A GeV/c at 
the JINR Nuclotron. An example of the central interaction of relativistic $^{11}$B in emulsion is given
 in Fig.~\ref{fig:111}. The $^{11}$B beam was directed parallel to the long side of the emulsion plane. Interactions 
were sought by viewing along the primary nucleus track. Over the total viewed -track length of 7141.5 cm we found
 542 interactions of $^{11}$B  with emulsion nuclei used in this analysis. In such a way, the mean free path was found to be 
$\lambda$=(13.2$\pm$0.6) cm. This value agrees well with the calculations by the geometric model. \par
\begin{figure*}
    \includegraphics[width=15cm]{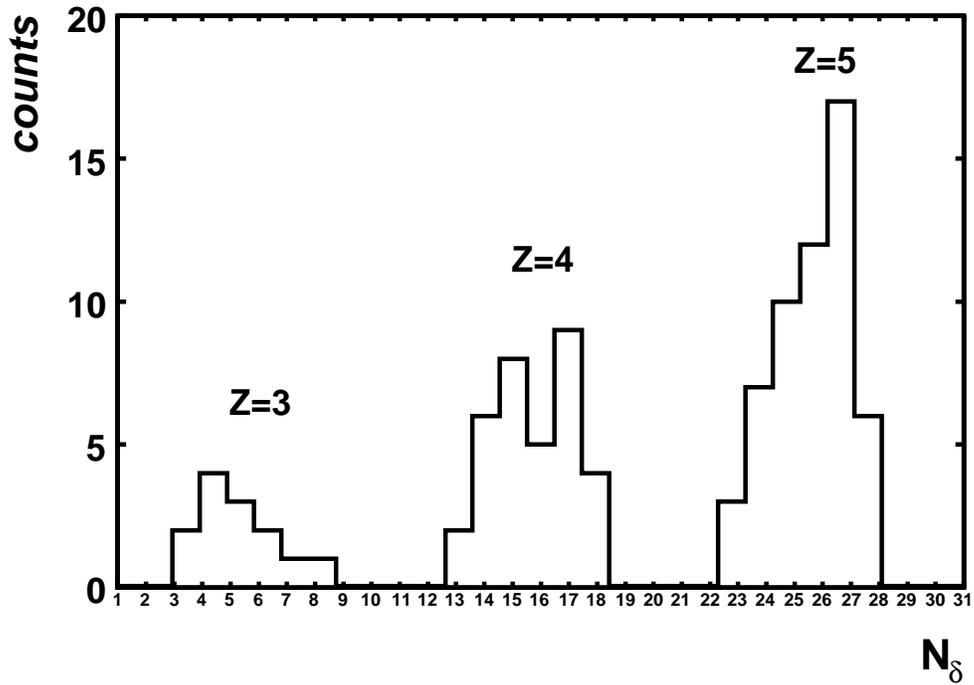}
    \caption{\label{fig:1} Distribution of number of $\delta$ electrons per 100 $\mu$m length on the tracks of
 relativistic fragments with charges Z$_{fr}$=3, 4, and 5.}
    \end{figure*}

\indent The relativistic fragment charge was determined by the method of counting the number of $\delta$ electrons on
the fragment track. The results of the determination of the charges Z$_{fr}$=3-5 by this method are given in Fig.~\ref{fig:1},
 which illustrates its high reliability.\par

\begin{figure*}
    \includegraphics[width=15cm]{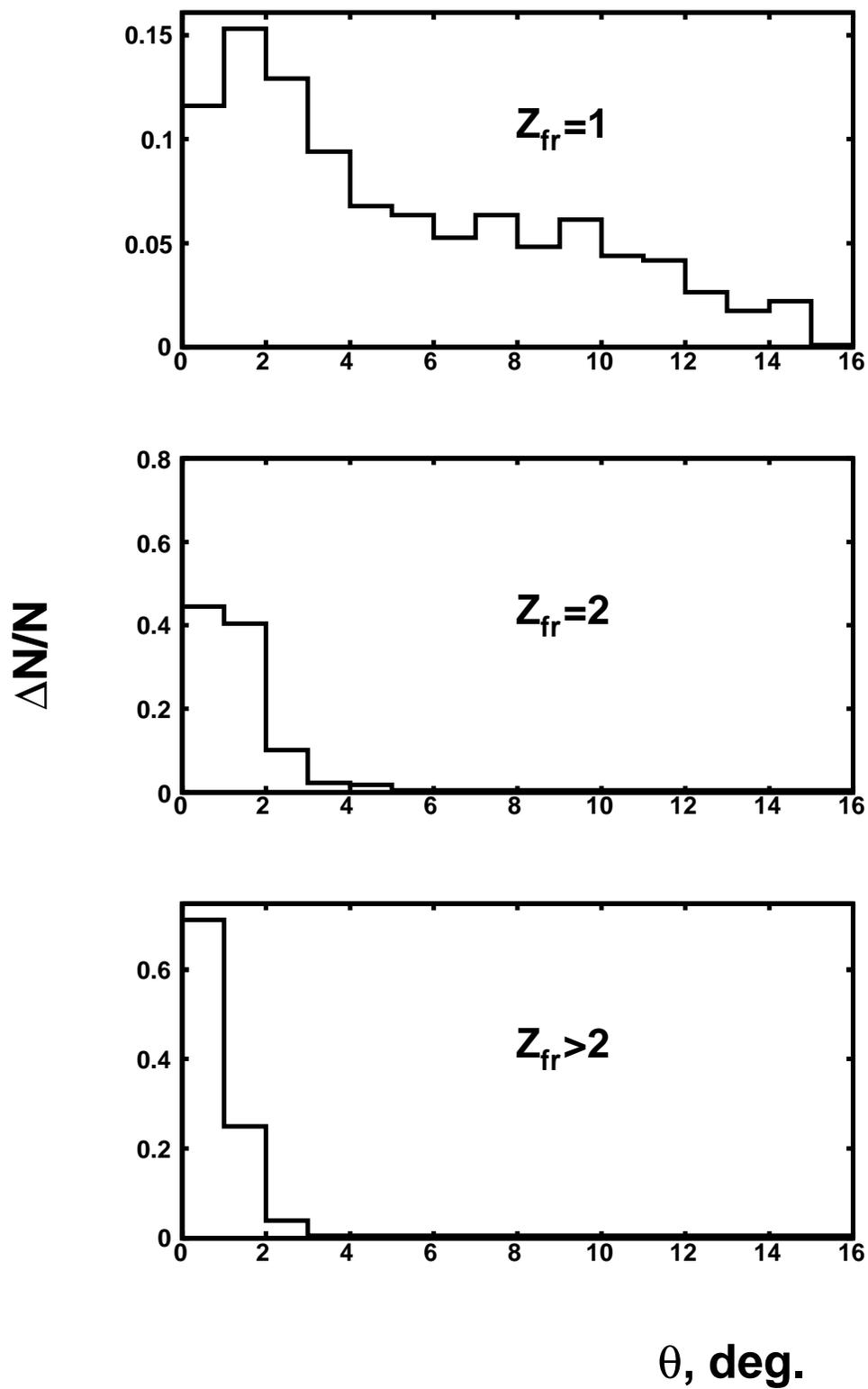}
    \caption{\label{fig:2} Normalized angular distributions of fragments with charges Z$_{fr}$=1 (above),Z$_{fr}$=2 (in the middle) and
Z$_{fr}>$2 (below).}
    \end{figure*}

 \indent The angular distributions of the $^{11}$B fragments are presented in Fig.~\ref{fig:2}
 separately for singly, doubly, and  multiply, Z$_{fr}>$2 charged fragments. The emission angles for Z$_{fr}>$2
are restricted in interval $\theta <$3$^\circ$, and the ones for doubly charged fragments to an interval $<$5$^\circ$.
The angles for singly charged particles were measured in an interval $<$15$^\circ$. Fig.~\ref{fig:2} (above) shows that the angular
 distribution changes its shape at about $\theta$=6$^\circ$. This shape of the angular distribution may be due to 
the fact that the singly charged particles are a mixture of the particles of the two kinds: relativistic  hydrogen 
isotopes and produced mesons the angular distributions of which are displaced relatively to one another. As momentum
measurements show the distribution for singly charged fragments occupy a region $\theta <$6-8$^\circ$. Basing on the
momentum measurements, the angular distribution shape and the estimation of the projectile fragmentation angle by the
equation 

\centerline{$sin \theta_{fr}=0.2/p_0=0.073 \rightarrow \theta_{fr}=4.16^{\circ}$}

a limiting angle for singly charged fragments was chosen to be $\theta$=6$^\circ$.\par
  
\section{\label{sec:level3}$^{11}$B Clustering}
\indent 
In order to study $^{11}$B cluster degrees of freedom use was made of the events in which the total charge of particles emitted within 
the fragmenting cone is equal to the charge of the projectile nucleus Q=$\Sigma$Z$_{fr}$. Such events were
divided into two classes A and B. Class A implies the breakup of a projectile not accompanied 
by the production of new particles, n$_s$=0. In their turn, the A class events can be subdivided into two groups: 
interactions without breakup of a target, n$_h$=0 (n$_h$=n$_b$+n$_g$) or \lq\lq white\rq\rq stars, and interactions
accompanied by the breakup of a target, n$_b<$8, n$_g$=0 in which the presence of several low-energy fragments is allowed.
These group which posess close characteristics are united into common class in order to increase statistics. Events of the class
A are notable for a low energy transfered to the projectile which results mostly in violations of intrinsic intercluster bonds.
Therefore  they are most interesting for the study of nuclear clustering. 
\par
\indent Class B implies peripheral but violent interactions of nuclei. In the events of this class, there can exist newly produced particles
 with emission angles $\theta >$15$^\circ$, as well as any quantity of the target fragments, n$_h>$0. The fragmentation channels for
 both classes are given in Table \ref{tab:1}.
\begin{table}
\caption{\label{tab:1} The charge topology distribution of the number of  interactions of the classes A and B, 
N$_{A}$ and  N$_{B}$, which were detected in an emulsion exposed to a $^{11}$B nucleus beam. Here  N$_{Z}$ is 
the number of fragments with charge Z. Numbers of \lq\lq white\rq\rq stars are shown in brackets. }

\begin{tabular}{c|c|c|c|c|c|c|c|c}
\hline\noalign{\smallskip}
\hline\noalign{\smallskip}

~~N$_{5}$~~&~~N$_{4}$~~&~~N$_{3}$~~&~~N$_{2}$~~&~~N$_{1}$~~&~~N$_A$~~&~~A, \% ~~&~~N$_B$~~&~~B, \% \\
~~1	~~&~~-	~~&~~-	~~&~~-	~~&~~-	~~&~~1	~~&~~4.6	~~&~~1	~~&~~1.7\\
~~-	~~&~~1	~~&~~-	~~&~~-	~~&~~1	~~&~~2	~~&~~9.4	~~&~~9	~~&~~15\\
~~-	~~&~~-	~~&~~1	~~&~~1	~~&~~-	~~&~~0	~~&~~0	~~&~~3	~~&~~5\\
~~-	~~&~~-	~~&~~1	~~&~~-	~~&~~2	~~&~~0	~~&~~0	~~&~~5	~~&~~8.3\\
~~-	~~&~~-	~~&~~-	~~&~~1	~~&~~3	~~&~~5	(1) ~~&~~24	~~&~~12	~~&~~20\\
~~-	~~&~~-	~~&~~-	~~&~~2	~~&~~1	~~&~~13	(6) ~~&~~62	~~&~~30	~~&~~52\\
~~-	~~&~~-	~~&~~-	~~&~~-	~~&~~5	~~&~~0	~~&~~0	~~&~~0	~~&~~0\\

\hline\noalign{\smallskip}
\hline\noalign{\smallskip}
\end{tabular}
%\hspace*{10cm}  % with the correct table height
\end{table}

\par 
\indent We can note some features in the data of Table \ref{tab:1} for events of classes A and B.
\begin{enumerate}
\item In both classes, the major fragmentation channel is $\Sigma$Z$_{fr}$=2+2+1: 62 and 52\%, respectively. For sake of
comparison, this channel for \lq\lq white\rq\rq stars produced in fragmentation of $^{10}$B nuclei amounts to 65\%.
\item Only 14\% of the events of the projectile breakup have fragments with charge Z$_{fr}>$2 while in peripheral
 interactions such events amount to 30\%.
\item In the projectile breakup events (A) there was observed no Li fragments, while in peripheral interactions (B)
 such events constitute 13\%.
\end{enumerate}
\par 
\begin{figure*}
    \includegraphics[width=15cm]{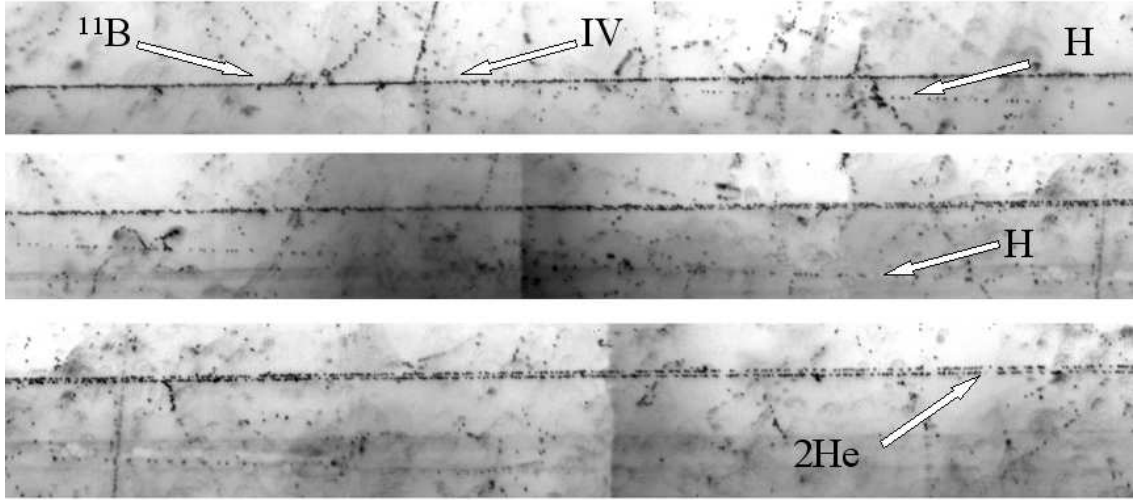}
    \caption{\label{fig:aba11} Example of peripheral interaction of a 2.75~A~GeV/c $^{11}$B nucleus in a nuclear track emulsion.
     The interaction vertex (indicated as {\bf IV}), nuclear fragment tracks ({\bf H} and {\bf He}) in a narrow angular
     cone, and fragment of target nucleus are seen  on the upper microphotograph. 
     Following the direction of the fragment jet, it is possible to distinguish 1 singly and 2 doubly charged 
     fragments on the middle and bottom microphotograph.}
    \end{figure*}
\begin{figure*}
    \includegraphics[width=15cm]{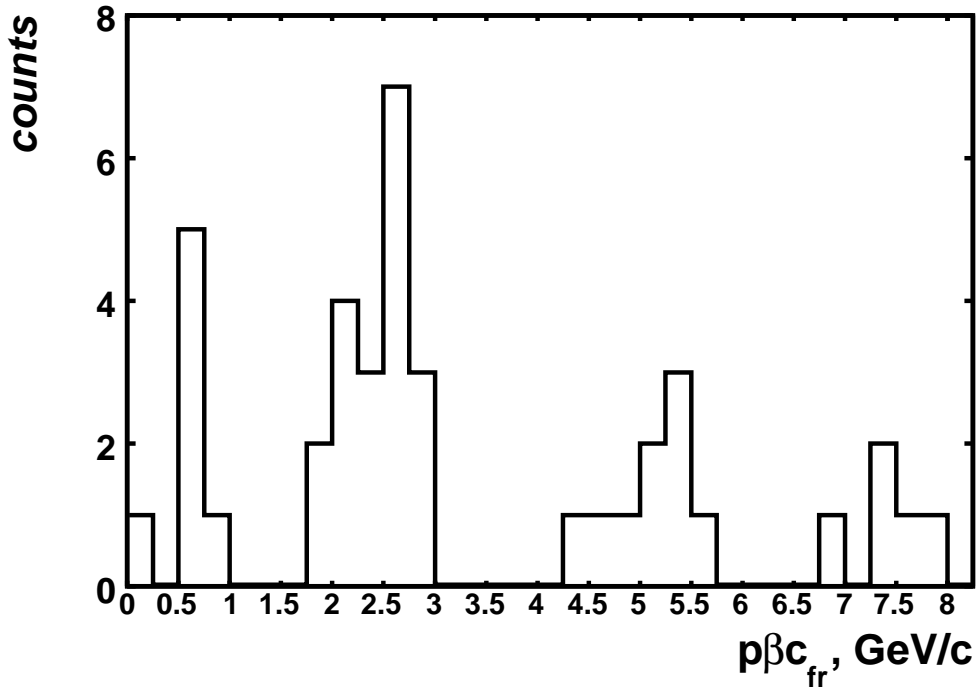}
    \caption{\label{fig:3} Momentum distribution for singly charged fragments.}
    \end{figure*}
\indent
The consimilar topology of \lq\lq white\rq\rq ~stars was investigated 
for $^{10}$B nuclei at the energy of 1.0 GeV per nucleon \cite{Adamovich04,Andreeva05}. 
 The fraction of the $^{10}$B$^*\rightarrow$d+$\alpha+\alpha$ decays is 40\% of the events with a
 charge topology 2+2+1. The contribution of the $^{10}$B$^*\rightarrow$d+$^{8}$Be+d$\rightarrow\alpha+\alpha$+d channel  
is estimated to be 18$\pm$3\%.  The probability of observing a 4+1 topology in the
 $^{10}$B$^*\rightarrow$p+$^{9}$Be decay is found small (3\%). Even being limited with current statistics one may conclude that the  3-body decays with
 a charge configuration  2+2+1 Fig.~\ref{fig:aba11})play a leading role in peripheral breakups of $^{11}$B as well as $^{10}$B nuclei in spite
of higher threshold values with the respect to the Li+He mode ones. It may be noticed, that
 the both breakup patterns  is indicative of an analogy  with the 3-body dissociation $^{12}$C$^*\rightarrow$3$\alpha$
 with one of $\alpha$ particles substituted by a deuteron or triton
\cite{Belaga95}. 
\par

\section{\label{sec:level4}Isotopic composition of singly charged fragments}
\indent
In order to study the basic $^{11}$B  fragmentation chahhel, $\Sigma$Z$_{fr}$=2+2+1 the singly charged fragment
 momenta were measured by the Coulomb multiple scattering method. The measurements enabled us to divide
 the singly charged fragments into protons, deuterons and tritons using the fact that the spectator fragments conserve
 the momentum per nucleon equal the primary one: $A_{fr}=(p\beta c)_{fr}/p_0$. The results of measurements are given in
 Fig.~\ref{fig:3}. As is seen, this method makes it possible to separate reliably the singly
 charged fragments by their mass.
\par

\indent
Thus we had determined for this channel the ratio between protons, deuterons and tritons: 
N$_p$:N$_d$:N$_t$=4:4:4 for $^{11}$B breakup events and N$_p$:N$_d$:N$_t$=17:5:1 for the $^{11}$B
 peripheral interactions (Table \ref{tab:2}).
\par 
\begin{table}
\caption{\label{tab:2} Isotopic composition of singly charged fragments for events with $\Sigma$Z$_{fr}$=2+2+1
of the classes A and B, 
N$_{A}$ and  N$_{B}$. }

\begin{tabular}{c|c|c|c|c|c|c}
\hline\noalign{\smallskip}
\hline\noalign{\smallskip}

~~ ~~&~~p~~&~~d~~&~~t~~&~~$\pi$~~&~~$\Sigma$~~\\
~~N$_A$~~&~~4	~~&~~4	~~&~~4	~~&~~0	~~&~~13\\
~~A, \%	~~&~~ 33	~~&~~ 33	~~&~~ 33	~~&~~0	~~&~~ \\
~~N$_B$	~~&~~17	~~&~~5	~~&~~1	~~&~~7	~~&~~30\\
~~B, \%	~~&~~ 57	~~&~~ 17	~~&~~3	~~&~~23	~~&~~	~~&~~ \\

\hline\noalign{\smallskip}
\hline\noalign{\smallskip}
\end{tabular}
%\hspace*{10cm}  % with the correct table height
\end{table}
\indent
\indent 
That is, it is possible to notice an essential decrease in the deuteron fraction and a practical disappearence of triton
when passing on from breakups to interactions. A large part of fragment tritons in $^{11}$B breakups
(about 1/3) testifies in favor of their existance as clusters weak internal bonds in $^{11}$B which are easily 
get destroyed in violent interaction processes. Besides, in 7 peripheral interactions (23\%) the momentum of singly charged 
particles was less than 1 GeV/c, that is, they may be either newly produced particles, or scattered protons. Thus, it is confirmed 
that the breakups of nuclei are more effective for the study of their cluster structure as compared with the violent interactions 
of nuclei.
\par 
\section{\label{sec:level5}Charge exchange of $^{11}$B to $^{11}$C$^*$}
\begin{figure*}
    \includegraphics[width=15cm]{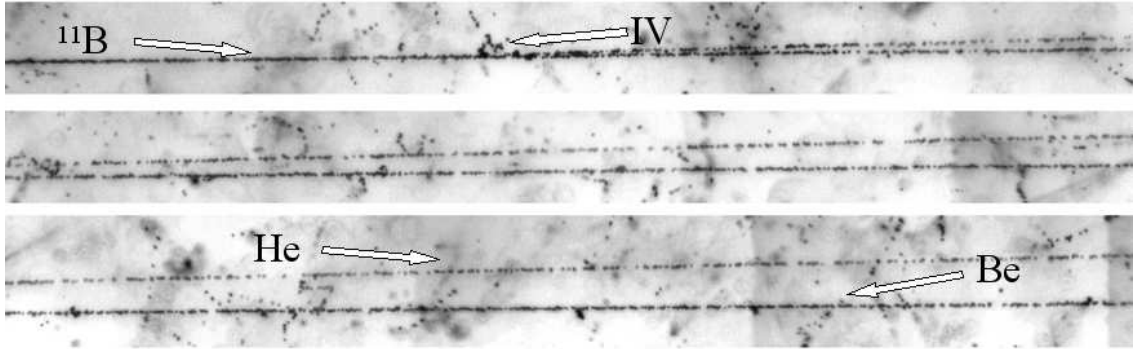}
    \caption{\label{fig:12} Example of the charge exchange of $^{11}$B $\rightarrow$ $^{11}$C$^*$ interaction  in a nuclear track emulsion.
     The interaction vertex (indicated as {\bf IV}), nuclear fragment tracks ({\bf He} and {\bf Be}) in a narrow angular
     cone are seen on the microphotographs. }
    \end{figure*}
\indent
The events of the classes A and B in which the charge of the primary track was 5 and total charge in the fragmentation
cone was $\Sigma$Z$_{fr}$=6 were ascribed to the inelastic charge exchange events above a particle decay threshold of
an excited $^{11}$C$^*$ nucleus. The statistics of the events of an inelastic charge 
exchange of $^{11}$B to $^{11}$C$^*$ is given in Table \ref{tab:3}. The particular feature  of the present experiment was the
observation of 8 events which can be considered as an inelastic charge exchange of $^{11}$B to $^{11}$C$^*$ 
followed by a breakup into 
two fragments with charges Z$_{fr}$=4 and Z$_{fr}$=2 Fig.~\ref{fig:12}). In order to avoid errors the charges in these events were measured several times.
Of these, 6 events belong to class A, in 5 events there is  none of the particles but the above mentioned fragments, and two events
belong to the class B. The fraction of such events is about 1.5\% of all events found in the initial scanning of the 
interactions, that is, the mean free path for the charge exchange of $^{11}$B to $^{11}$C$^*$ $\lambda$ is equal to
(0.89$\pm$0.32) m. 
\par 
\begin{table}
\caption{\label{tab:3} Fragmentation channels for events with charge exchange of $^{11}$B to $^{11}$C$^*$
of the classes A and B, 
N$_{A}$ and  N$_{B}$. Here  N$_{Z}$ is 
the number of fragments with charge Z.}

\begin{tabular}{c|c|c|c|c|c|c}
\hline\noalign{\smallskip}
\hline\noalign{\smallskip}

~~N$_{5}$~~&~~N$_{4}$~~&~~N$_{3}$~~&~~N$_{2}$~~&~~N$_{1}$~~&~~N$_A$~~&~~N$_B$~~\\
~~1	~~&~~-	~~&~~-	~~&~~-	~~&~~1	~~&~~0	~~&~~1\\
~~-	~~&~~1	~~&~~-	~~&~~1	~~&~~-	~~&~~6	~~&~~2\\
~~-	~~&~~1	~~&~~-	~~&~~-	~~&~~2	~~&~~0	~~&~~7\\
~~-	~~&~~-	~~&~~1	~~&~~-	~~&~~3	~~&~~0	~~&~~2\\
~~-	~~&~~-	~~&~~-	~~&~~2	~~&~~2	~~&~~0	~~&~~3\\

\hline\noalign{\smallskip}
\hline\noalign{\smallskip}
\end{tabular}
%\hspace*{10cm}  % with the correct table height
\end{table}
\indent Table \ref{tab:3} clearly demonstrate appearance of low-lying cluster modes when collision inelasticity has
a minimal obsevability.
The decay of $^{11}$C nucleus into two particles with charges 4 and 2 can occur only in the
 $^{11}$C$\rightarrow^{7}$Be+$^{4}$He. In the events of the class A, there was found no inelastic 
charge exchange of $^{11}$B to $^{11}$C with a decay in other channels. However, in order to establish unabiguously
the fact of the charge exchange it is required to carry out He momentum measurements.
Nevertheless, in distinction to the considered above $^{11}$B case in Table \ref{tab:1} there is indication in on more important role of 2-body dissociations of excited 
 $^{11}$C than 3-body ones - A class events with final state 3He were not observed (Table \ref{tab:3}). 
This obvious difference may be originated due to a higher Coulomb barrier in a $^{11}$C nucleus.
This feature has to be verified in special $^{11}$C study provided with emulsions exposured in a secondary beam of these nuclei
produced in the charge exchange process $^{11}$B$\rightarrow^{11}$C. 

\par 
\section{\label{sec:level6}Summary}
\indent This paper provides a framework for a longer time-demanding analysis. Already  found $^{11}$B collisions
will allow us to improve our statistics by factor 4-5 to strengthen the conlclusions.
Nevertheless, in spite of rather limited statistics analyzed to the present time one can derive the important 
features of peripheral fragmentation
of $^{11}$B nuclei. Firstly, 3-cluster dissociations H+2He provide a leading contribution to the breakup cross-section.
Secondly, there is an indication on a strong triton contribution comparable  with $^{7}$Li$^*\rightarrow\alpha+t$.
Thirdly, a $^{11}$B nucleus fragmentation is promising source to populate $^{11}$C$^*$ excited states in charge exchange processes
directly or via special formation of the $^{11}$C secondary beam in a charge exchange process. 
\par
\begin{acknowledgments}
\indent The work was supported by the Russian Foundation for Basic Research
 (grant 04-02-16593),
 VEGA N1/2007/05.  Grant from the Agency for Science of the Ministry for Education of the
 Slovak Republic and the Slovak Academy of Sciences, and Grants from the JINR
 Plenipotentiaries of Bulgaria, the Slovak Republic, the Czech Republic
 and Romania in the years 2002-2006.\par
  
\end{acknowledgments}

\end{document}